\documentclass[aps,prb,amsfonts,amssymb,twocolumn,amsmath,preprintnumbers,nofootinbib,floatfix,
showpacs]{revtex4-1}
\usepackage[dvips]{graphics}
\usepackage{graphicx}
\usepackage{bm}
\begin{document}

\title{Triplet contribution to the Josephson current in the 
nonequilibrium superconductor/ferromagnet/superconductor junction}
\author{I. V. Bobkova}
\affiliation{Institute of Solid State Physics, Chernogolovka,
Moscow reg., 142432 Russia}
\author{A. M. Bobkov}
\affiliation{Institute of Solid State Physics, Chernogolovka,
Moscow reg., 142432 Russia}

\date{\today}

\begin{abstract}
The Josephson current through a long s-wave superconductor/weak ferromagnet/s-wave superconductor
weak link is studied theoretically in the regime of nonequilibrium spin-dependent occupation
of electron states in the ferromagnetic intelayer. While under the considered nonequilibrium
conditions the standard supercurrent, carried by the singlet part of current-carrying density of states, practically is not
modified, the additional supercurrent flowing via the triplet part of the current-carrying density of states
appears. Depending on voltage, controlling the particular form of spin-dependent 
nonequilibrium in the interlayer, this additional current can enhance or reduce the usual current of the singlet
component and also switch the junction between $0$- and $\pi$-states.  
\end{abstract}

\maketitle

New states have been predicted and observed in Josephson weak links in the last years. One of them has common
origin with the famous LOFF-state \cite{larkin64,fulde64}. This mesoscopic LOFF-state, which is induced 
in superconductor/weak ferromagnet/superconductor (SFS) Josephson junction, was predicted 
theoretically \cite{buzdin82,buzdin92} and observed experimentally \cite{ryazanov01,kontos02,blum02,guichard03}.
In this state Cooper pair acquires the total momentum
$2Q$ or $-2Q$ inside the ferromagnet as a response
to the energy difference between the two spin directions. Here $Q \propto h/v_f$, 
where $h$ is an exchange energy and $v_f$ is
the Fermi velocity. Combination of the two possibilities
results in the spatial oscillations of the condensate wave function $\Psi (x)$ in the ferromagnet along the direction
normal to the SF interface. $\Psi_{s} (x) \propto \cos(2Qx)$ for the singlet Cooper pair \cite{demler97}. The same
picture is also valid in the diffusive limit. The only thing we need to add is an extra decay of the condensate
wave function due to scattering. In the regime $h \gg |\Delta|$, where $\Delta$ is a superconducting order parameter
(OP) in the leads, the decay length is equal to the magnetic coherence length $\xi_F=\sqrt{D/h}$, while 
the oscillation period is given by
$2\pi \xi_F$. Here $D$ is the diffusion constant in the ferromagnet, $\hbar=1$ throughout the paper.

The presence of an exchange field also leads to
the formation of the triplet component of the condensate wave
function in the interlayer. In the case of a homogeneous exchange
field only the component with zero spin projection on the field direction $S_z =
0$ is induced. Combining the two pairs with the total momenta $2Q$
and $-2Q$ into the triplet combination we see that the in-plane
($S_z = 0$) triplet condensate wave function component $\Psi_{t}
(x) \propto \sin(2Qx)$, that is oscillates in space with the same
period as the singlet one, but is shifted by $\pi/2$ with respect
to it. Here we do not discuss the other triplet components with
$S_z=\pm 1$, which are typically induced in case of inhomogeneous
magnetization \cite{bergeret05}.

The energy spectrum of the superconducting correlations is expressed in a so-called supercurrent-
carrying density of states (SCDOS) \cite{volkov95,wilheim98,yip98, heikkila02}. 
This quantity represents the density of states
weighted by a factor proportional to the current that each state carries in a certain direction. 
Under equilibrium conditions the supercurrent can be expressed via the SCDOS as \cite{yip98}
\begin{equation}
j \propto \int d \varepsilon N_j (\varepsilon)\tanh \varepsilon/2T
\label{supercurrent}
\enspace ,
\end{equation}
where $\varepsilon$ stands for the quasiparticle energy,
$\varphi(\varepsilon)=\tanh \varepsilon/2T$ is the equilibrium distribution function and $N_j(\varepsilon)$ is SCDOS.
In the presence of spin effects SCDOS becomes a matrix $2 \times 2$ in spin space and can be represented as 
$\hat N_j=N_{j,s}+\bm N_{j,t}\bm \sigma$, where $\sigma_i$ are Pauli matrices in spin space.   
The scalar in spin space part of SCDOS $N_{j,s}$ is referred to as the singlet part of SCDOS in the paper
and the vector part $\bm N_{j,t}$ is referred to as the triplet part. $\bm N_{j,t}$ is directly proportional
to the triplet part of the condensate wave function. It is well known that the spin supercurrent 
can not flow through the singlet superconducting leads. Therefore, $\bm N_{j,t}$ 
does not contribute to the supercurrent in equilibrium. Having in mind that the triplet part of SCDOS 
is even function of quasiparticle energy, one can directly see that this is indeed the case.  

Under nonequilibrium conditions one can change the value of the critical Josephson current through the junction
and even realize the $\pi$-state by manipulating the quasiparticle distribution in an interlayer region.
This effect was predicted theoretically \cite{volkov95,wilheim98} and observed experimentally 
\cite{baselmans99,huang02,crosser08} for a diffusive  SNS junction. The point is that positive and negative parts of
SCDOS give energy-dependent contributions
to the supercurrent in the positive and negative direction. The size and direction of the
total supercurrent depends therefore on the occupied fraction of
these states, which is analogous to the occupation of the discrete
Andreev bound states in a ballistic system. That is one can  obtain negative Josephson current response
to small phase differences and, hence, switch the system into $\pi$-state by creating an appropriate nonequilibrium
quasiparticle distribution in the weak link region. The combination of the exchange field $h \ll |\Delta|$
and the spin independent nonequilibrium distribution function has been considered as well \cite{heikkila00}. 
Under these conditions the influence of the nonequilibrium distribution function is also
consists in the redistribution of the quasiparticles between the energy levels. However, it was shown that 
in this limit of small exchange fields the combined effect of the exhange field and the nonequilibrium 
distribution function is nontrivial. For instance, part of the field-suppressed supercurrent can be recovered
by adjusting a voltage between additional electrodes, which controls the ditribution function.      

In the present paper we investigate the effect of nonequilibrium occupation of the supercurrent-carrying states
on the Josephson current in SFS junction in the parameter range $|\Delta| \ll h \ll \varepsilon_f$, where $\varepsilon_f$
is the Fermi energy of the ferromagnet. This regime is relevant to weak ferromagnetic alloys, which were used for
the experimental realization of magnetic $\pi$-junctions. It is shown
that if the distribution function becomes nonequilibrium and spin-dependent, the supercurrent carried by
the SCDOS triplet component $\bm N_{j,t}$ in the ferromagnet is non-zero. 
The magnitude of this current cotribution $j_{t}$ can be of the same order or even larger than the current 
contribution $j_s$ carried by the singlet component $N_{j,s}$. Due to the fact that the singlet and triplet
components of the anomalous Green's function have the same oscillation period but shifted in phase by $\pi/2$, 
$j_{t}$ can increase the usual supercurrent,
carried by the singlet component of SCDOS, or weaken it, or even reverse the sign of the total supercurrent, thus switching
between $0$ and $\pi$-state. Experimentally the most probable way to realize the spin-dependent nonequilibrium
in the interlayer is to apply a voltage to (or to pass the dissipative current through) a spin-active material. Then
under appropriate conditions even quite small voltages should be enough to switch the system from $0$ to $\pi$-state
and vice versa. 

That is we consider another mechanism of supercurrent manipulation by creating a nonequilibrium
quasiparticle distribution in the interlayer, which cannot be reduced to the redistribution of the quasiparticles
between the energy levels. In principle, the both mechanisms can be realized in a junction simultaneously.
However, in this particular study we assume $h \gg \Delta$ and the Thouless energy
$\varepsilon_{Th}=D/d^2 \ll h$, that is the interlayer length $d \gg \xi_F$.
As it is shown below, in this regime (and for the case of low-transparency SF interfaces) a spin-independent 
nonequilibrium distribution of quasiparticles
in the ferromagnet practically does not affect the Josephson current, that is the described above  mechanism
of the critical current reversal by the spin-independent redistribution of supercurrent-carrying states population
is irrelevant in this case. For high-transparency SF interfaces the both mechanisms contribute to supercurrent.

For a quantitative analysis we use the formalism of quasiclassical Green-Keldysh functions in the diffusive limit
\cite{usadel}.
The fundamental
quantity for diffusive transport is the momentum average of the
quasiclassical Green's function $\check g(x,\varepsilon) =
\langle \check g(\bm p_f, x,\varepsilon) \rangle_{\bm p_f}$. It
is a $8\times8$ matrix form in the product space of Keldysh,
particle-hole and spin variables. Here $x$ is the coordinate
measured along the normal to the junction.

The electric current should be calculated via Keldysh part of the
quasiclassical Green's function. For the plane diffusive junction
the corresponding expression reads as follows
\begin{equation}
j = \frac{-d}{e R_F} \int
\limits_{-\infty}^{+\infty} \frac{d \varepsilon}{8 \pi^2} {\rm Tr}_4
\left[\frac {\tau_0 + \tau_3}{2}
\left(\check g(x, \varepsilon)\partial _x \check
g(x, \varepsilon)\right)^K \right] \label{tok}
,
\end{equation}
where $e$ is the electron charge and $R_F$ stands for the resistance of the ferromagnetic region.
$\left(\check g(x, \varepsilon)\partial_x \check
g(x, \varepsilon)\right)^K$ is  $4\times4$
Keldysh part of the corresponding combination of full Green's
functions. $\tau_i$ are Pauli matrices in
particle-hole space.

It is convenient to express Keldysh part of the full Green's function via the retarded and advanced components
and the distribution function: $\check g^K=\check g^R \check \varphi-\check \varphi \check g^A$. Here the argument
$(x, \varepsilon)$ of all the functions is omitted for brevity. The distribution function is diagonal in particle-hole
space: $\check \varphi=\hat \varphi (\tau_0+\tau_3)/2+ \sigma_2 \hat {\tilde \varphi} \sigma_2
(\tau_0-\tau_3)/2$. All the matrices denoted by $\hat ~$ are $2 \times 2$ matrices in spin space throughout
the paper. In terms of the distribution function current (\ref{tok}) takes the form

\begin{widetext}
\begin{equation}
j = \frac{-d}{e R_F} \int
\limits_{-\infty}^{+\infty} \frac{d \varepsilon}{8 \pi^2} {\rm Tr}_2
\left[-\pi^2 \partial _x \hat \varphi - \hat
g^R \partial _x \hat \varphi \hat g^A -  
\hat f^R \partial_x \hat {\tilde \varphi}\hat {\tilde f}^A +
(\hat g^R \partial_x \hat g^R +\hat f^R \partial_x \hat {\tilde f}^R) \hat \varphi- 
\hat \varphi (\hat g^A \partial_x \hat g^A +\hat f^A \partial_x \hat {\tilde f}^A) \right]
\label{current_distribution}
.
\end{equation}
\end{widetext}
We assume that the direction of the exchange field $\bm h$ is spatially homogeneous and choose the quantization axis
along the field. In this case the distribution
function and the normal part $\hat g^{R,A}$ of the Green's function are diagonal matrices in spin space. 
The anomalous Green's functions can be represented as $\hat f^{R,A}=\hat f_d^{R,A}i \sigma_2$ and 
$\hat {\tilde f}^{R,A}=-i\sigma_2\hat {\tilde f}_d^{R,A}$, where $\hat f_d^{R,A}$ and 
$\hat {\tilde f}_d^{R,A}$ are diagonal in spin space.

The retarded and advanced Green's functions are obtained by solving the Usadel equations \cite{usadel}
supplemented with Kupriyanov-Lukichev boundary conditions at SF interfaces \cite{kupriyanov88}. It is worth
to note that we can safely apply these boundary conditions to the problem
of plane diffusive junction even for high enough dimensionless conductance $g$ of the SF interface. 
This can be done in spite of the fact that they are the linear in transparency  approximation of more general
Nazarov boundary conditions \cite{nazarov99}. The reason is that the effective number of interface channels $N \sim d_y/l$ 
is large and a separate channel transparency $T \sim g(l/d_y)$ is usually considerably less than unity. 
Here $d_y$ is the junction width and $l$ is the mean free path. 

Further, the condition $d \gg \xi_F$ 
allows us to find the solution analytically
even for an arbitrary SF interface transparency and low temperature, that is in the parameter region, where the
equations cannot be linearized with respect to the anomalous Green's function. We start from
the completely incoherent junction (that is, consider the left and right SF interfaces separately) and then calculate
the corrections up to the first order of the small parameter $\exp{[-d/\xi_F]}$
to the Green's functions. Within this accuracy the anomalous Green's functions in the vicinity of left 
and right SF interfaces (at $x=\mp d/2$) take the form
\begin{eqnarray}
f_{d \sigma}^{R,A}=\kappa i \pi \left[ \sinh \Theta_{\sigma}^{R,A}e^{-i\alpha \chi/2}+
4 \Sigma_\sigma^{R,A}(x) e^{i\alpha \chi /2} \right]
\enspace , \nonumber \\
\tilde f_{d \sigma}^{R,A}=-f_{d \sigma}^{R,A}(\chi \to -\chi)\enspace .
\qquad \qquad
\label{f}
\end{eqnarray}
Here $\sigma=\uparrow ,\downarrow $ (or $+1(-1)$ within Equations) is the electron spin projection on 
the quantization axis, $\kappa=+1(-1)$ corresponds
to the retarded (advanced) functions, $\alpha =+1(-1)$ in the vicinity of the left (right) SF interface and
$\chi$ is the order parameter phase difference between the superconducting leads. 
The first term represents the anomalous Green's function at the ferromagnetic side of the isolated SF boundary
and does not enter the following results. So, we do not give it explicitly.
The second term is the first order correction, originated from the anomalous Green's function
extended from the other SF interface:
\begin{equation}
\Sigma _\sigma^{R,A}(x)=K_\sigma ^{R,A}e^{\displaystyle -(\frac{d}{2}-\alpha x)(1+i\kappa \sigma)/\xi_F}
\label{Sigma}
\enspace ,
\end{equation}
where $K_\sigma^{R,A}$ is determined by the equation
\begin{eqnarray}
(1+i\kappa \sigma)K_\sigma^{R,A}(1-{K_\sigma^{R,A}}^2)=
\frac{R_F \xi_F}{4R_g d}\left[ \sinh \Theta_s^{R,A}(1+ \right. \nonumber \\
\left. +{K_\sigma^{R,A}}^2+{K_\sigma^{R,A}}^4)-
\cosh \Theta_s^{R,A} 4 K_\sigma^{R,A}(1+{K_\sigma^{R,A}}^2)\right]
\label{K}
\enspace .
\end{eqnarray}
$R_g$ stands for the resistance of each SF interface, which are
supposed to be identical. $\sinh \Theta_s^{R,A}$ and $\cosh
\Theta_s^{R,A}$ originate from the Green's functions at the 
superconducting side of SF interfaces. We
assume that the parameter $(R_F \xi_s/R_g d)(\sigma_F/\sigma_s) \ll 1$, where
$\xi_s=\sqrt{D/\Delta}$ is the superconducting coherence length in the leads, 
$\sigma_F$ and $\sigma_s$ stand for conductivities of
ferromagnetic and superconducting materials, respectively. It 
allows us to neglect the suppression of the
superconducting order parameter in the S leads near the interface
and take the Green's functions at the superconducting side of the
boundaries to be equal to their bulk values. In this case
\begin{eqnarray}
\cosh \Theta_s^{R,A}=\frac{-\kappa
i\varepsilon}{\sqrt{|\Delta|^2-(\varepsilon+\kappa i\delta)^2}}
\nonumber \\
\sinh \Theta_s^{R,A}=\frac{-\kappa i
|\Delta|}{\sqrt{|\Delta|^2-(\varepsilon+\kappa i \delta)^2}}
\label{bulk_Greens_functions} \enspace ,
\end{eqnarray}
where $\delta$ is a positive infinitesimal.

The SCDOS, entering the current (\ref{current_distribution}) takes the form
\begin{eqnarray}
N_{j\sigma}(\varepsilon)=\left( \hat g^R \partial_x \hat g^R +\hat f^R \partial_x \hat {\tilde f}^R- 
\hat g^A \partial_x \hat g^A -\hat f^A \partial_x \hat {\tilde f}^A \right)_\sigma = 
\nonumber \\ 
=\frac{8\pi^2 \sin \chi}{\sigma_F R_g}{\rm Im}\left[ \Sigma_\sigma^R(x=-\alpha \frac{d}{2})\sinh 
\Theta_s^R \right]
\label{SCDOS}
\enspace .~~~~~~~~~~
\end{eqnarray}

The nonequilibrium distribution function in the interlayer is
proposed to be created by applying a voltage along the $y$
direction between two additional electrodes $N_b$ and $N_t$, which
are attached to the central part of the interlayer. It
is supposed that the conductances of $N_b$F and $N_t$F interfaces
are spin-dependent and equal to ${g_b}_\sigma$ and ${g_t}_\sigma$,
respectively. The voltage ${V_t}_\sigma$ (${V_b}_\sigma$) between the superconducting leads
and $N_t$ ($N_b$) electrode can also be spin-dependent. It can be realized, for example,
by attaching one (or both) of the electrodes $N_b$ or $N_t$ to a strong ferromagnet and applying a voltage 
between the other one and the ferromagnet.

In order to obtain the distribution function up to the first order of the parameter 
$\exp{[-d/\xi_F]}$ we solve the kinetic equation for the distribution function,
which is derived from the Keldysh part of the Usadel equation. The
boundary conditions to the kinetic equation are also obtained from
the Keldysh part of the general Kypriyanov-Lukichev boundary
conditions. Further it is assumed that 
$|{eV_{t,b}}_{\uparrow ,\downarrow}|<|\Delta|$. Under this condition
the part of the current associated with the first three terms in Eq.~(\ref{current_distribution})
is only determined by the first order correction $\Sigma_{\sigma}^{R,A}$ to the anomalous
Green's function and the distribution function $\varphi^{(0)}_\sigma$, calculated up to the zero
order of the parameter $\exp{[-d/\xi_F]}$:
\begin{equation}
\left(-\pi^2 \partial _x \hat \varphi - \hat
g^R \partial _x \hat \varphi \hat g^A -  
\hat f^R \partial_x \hat {\tilde \varphi}\hat {\tilde f}^A \right)_\sigma \!\!\!= 
N_{j\sigma}\frac{\tilde \varphi_\sigma^{(0)}-\varphi_\sigma^{(0)}}{2}, 
\label{diffusion}
\end{equation}
where $N_{j\sigma}(\varepsilon)$ is expressed by Eq.~(\ref{SCDOS}). 

The distribution function $\varphi^{(0)}$ does not depend on $x$.
So, it is convenient to calculate it in the middle of the interlayer, where disregarding the parameter
$\exp{[-d/\xi_F]}$ means disregarding the superconducting proximity effect. Then, by considering $N_b/F/N_t$
junction and applying Kypriyanov-Lukichev boundary conditions $N_b/F$ and $F/N_t$ interfaces, we come to
the following expression for $\varphi^{(0)}$ (inelastic scattering processes are not taken into account):   
\begin{eqnarray}
\varphi^{(0)}_\sigma=\frac{\tanh \frac{\varepsilon-e{V_t}_\sigma}{2T}{g_t}_\sigma(\sigma_F+d_y {g_b}_\sigma)
}{\sigma_F({g_t}_\sigma+{g_b}_\sigma)+2d_y{g_t}_\sigma{g_b}_\sigma}+ \nonumber \\
+\frac{\tanh \frac{\varepsilon-e{V_b}_\sigma}{2T}{g_b}_\sigma (\sigma_F+d_y {g_t}_\sigma)}
{\sigma_F({g_t}_\sigma+{g_b}_\sigma)+2d_y{g_t}_\sigma{g_b}_\sigma}
\label{phi_N}
\enspace ,
\end{eqnarray}
$\tilde {\varphi}^{(0)}_\sigma$ is
connected to $\varphi^{(0)}_\sigma$ by the symmetry relation $\tilde {\varphi}^{(0)}_{\uparrow ,\downarrow}(\varepsilon)=
-\varphi^{(0)}_{\downarrow ,\uparrow}(-\varepsilon)$. We focus on the case 
$g_{t\sigma} \ll \sigma_F/d_y$ or $g_{b\sigma} \ll \sigma_F/d_y$, when the $y$-dependence of the distribution 
function $\varphi^{(0)}$ can be disregarded. 

Substituting Eqs.~(\ref{SCDOS})-(\ref{diffusion}) into Eq.~(\ref{current_distribution}) 
we find that the Josephson current takes the form 
\begin{equation}
j=\frac{-d}{2eR_F}\int \frac{d \varepsilon}{8\pi^2} \sum \limits_\sigma 
\left[ (\varphi^{(0)}_\sigma+\tilde {\varphi}^{(0)}_\sigma) N_{j\sigma}\right]
\label{current_general}
\enspace ,
\end{equation}
where $N_{j\sigma}(\varepsilon)$ is expressed by Eq.~(\ref{SCDOS}) and $\varphi^{(0)}_\sigma$ should be taken from
Eq.~(\ref{phi_N}). Elecrtical current (11) can be divided into two parts:
\begin{eqnarray}
j=j_s+j_t \nonumber
\enspace ,~~~~~~~~~~~~~~~~~~~~\\
j_s=j_{s,c} \sin \chi=\frac{-d}{eR_F}\int \frac{d \varepsilon}{8\pi^2} 
\left[ (\varphi^{(0)}_0+\tilde {\varphi}^{(0)}_0) N_{j,s}\right] 
\nonumber
\enspace , \\
j_t=j_{t,c}\sin \chi =\frac{-d}{eR_F}\int \frac{d \varepsilon}{8\pi^2} 
\left[ (\varphi^{(0)}_z+\tilde {\varphi}^{(0)}_z) N_{j,t}\right]
\label{current_general_parts}
\enspace ,
\end{eqnarray}
where $\varphi_0=(\varphi_\uparrow +\varphi_\downarrow )/2$, $\varphi_z=(\varphi_\uparrow -\varphi_\downarrow )/2$,
$\tilde \varphi_{0,z}=(\tilde \varphi_\uparrow \pm \tilde \varphi_\downarrow )/2=\mp \varphi_{0,z}(-\varepsilon)$,
$N_{j,s}=(N_{j\uparrow}+N_{j\downarrow})/2$ is the singlet part of SCDOS and 
$N_{j,t}=(N_{j\uparrow}-N_{j\downarrow})/2$ is the z-component of the SCDOS triplet part (the other components
equal to zero for the considered case of homogeneous magnetization). It is seen from Eq.~(\ref{current_general_parts})
that $N_{j,t}$ gives rise to the additional contribution to the spinless electrical current if 
the quasiparticle distribution is spin-dependent.

\begin{figure*}[!tbh]
  \centerline{\includegraphics[clip=true,width=0.95\linewidth]{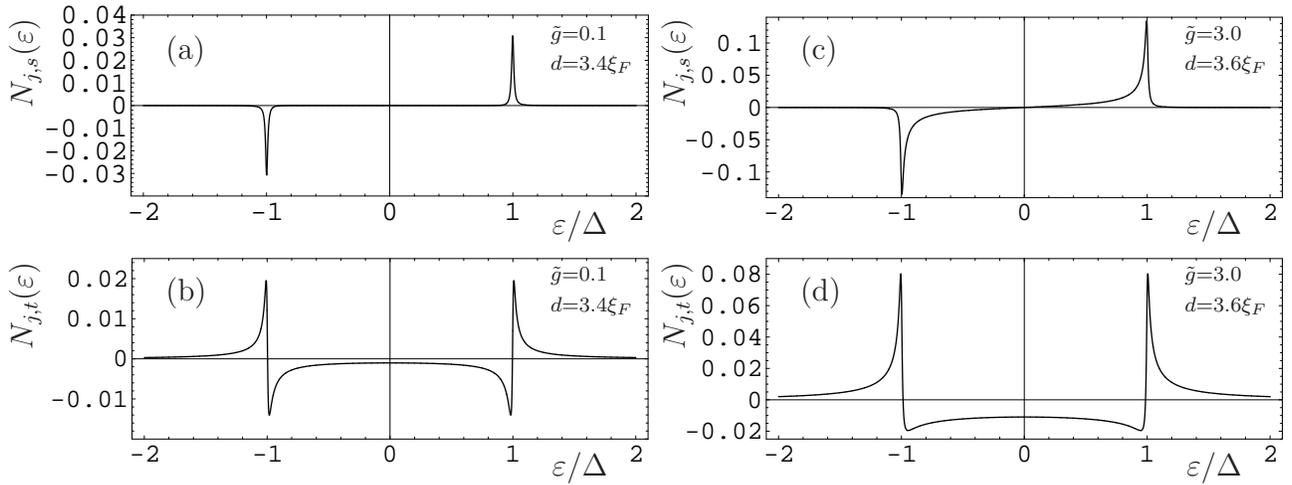}}
   \caption{The singlet and triplet parts of SCDOS in dependence on the quasiparticle energy.
The left column corresponds to the low-transparency limit $\tilde g=0.1$, while the right column
represents the high-transparency case $\tilde g=3.0$.}
\label{SCDOS_fig}
\end{figure*}

While Eqs.~(\ref{current_general})-(\ref{current_general_parts}) are valid for arbitrary SF interface
transparency, at first we concentrate on the discussion of the tunnel limit $\tilde g \equiv R_F\xi_F/R_gd \ll 1$, 
where Eq.~(\ref{K}) can be easily solved and the integral over energy can be calculated analytically. 
For the analytical analysis we choose the most simple form for the distribution function Eq.~(\ref{phi_N}) 
by setting ${g_t}_\sigma \to 0$ and ${V_b}_\downarrow = 0$. 
Then $\varphi^{(0)}_\uparrow = \tanh[(\varepsilon-e{V_b}_\uparrow)/2T]$, while 
$\varphi^{(0)}_\downarrow = \tanh[\varepsilon/2T]$. As it is demonstrated below, the results 
corresponding to another set of parameters, which determines the particular form of the distribution
function satisfying Eq.~(\ref{phi_N}), qualitatively do not differ from that ones represented here.
At low temperature $T \ll |eV_{b\uparrow}|$
we obtain for the Josephson current the following result     
\begin{eqnarray}
j=\frac{R_F \xi_F \sin \chi}{4eR_g^2 d}|\Delta| e^{-d/\xi_F}\left[ 
\sqrt 2 \pi \cos(\frac{d}{\xi_F}+\frac{\pi}{4})+\right. \nonumber \\
\left. \frac{1}{\sqrt 2}\log \left| \frac{|\Delta|+e{V_b}_\uparrow}{|\Delta|-e{V_b}_\uparrow} \right|
\sin (\frac{d}{\xi_F}+\frac{\pi}{4}) \right]
\label{current_tunnel}
\enspace .
\end{eqnarray}

The first term in Eq.~(\ref{current_tunnel}) represents the part of the supercurrent $j_s$ carried by the singlet
component of SCDOS. Under the conditions $T \ll |eV_{t,b}|$ and
$|{eV_{t,b}}_{\uparrow ,\downarrow}|<|\Delta|$ it is not affected by the fact that the distribution function 
is nonequilibrium, as can be seen from Eq.~(\ref{current_tunnel}). 
The reason is that for a long junction and $h \gg |\Delta|$ in the tunnel limit 
the singlet part of SCDOS is concentrated in the narrow energy intervals around $\varepsilon=\pm |\Delta|$, 
as it is illustrated in panel (a) of Fig.~\ref{SCDOS_fig}. This is opposed to the cases of diffusive
SNS junction with $h=0$ \cite{volkov95,wilheim98,yip98,heikkila02} and SFS junction with $h \ll \Delta$
\cite{heikkila00}, where the singlet part of SCDOS is finite
and exhibits nontrivial energy dependence in the subgap energy region $|\varepsilon|<|\Delta|$.
Under the conditions $T \ll |eV_{t,b}|$ and
$|{eV_{t,b}}_{\uparrow ,\downarrow}|<|\Delta|$ the distribution function 
$\varphi_0^{(0)}+\tilde \varphi_0^{(0)} \approx ({\rm sgn}[\varepsilon-eV_{b\uparrow}]+
{\rm sgn}[\varepsilon+eV_{b\uparrow}]+2{\rm sgn}[\varepsilon])/2$, 
which enters the expression for $j_s$ [Eq.(\ref{current_general_parts})],
practically does not differ from its equilibrium value for energy intervals around $\varepsilon=\pm |\Delta|$. That is,
under the considered conditions the widely discussed in the literature mechanism of supercurrent manipulation
by nonequilibrium redistribution of quasiparticles between energy levels 
\cite{volkov95,wilheim98,baselmans99,huang02,crosser08,heikkila00} does not contribute.
   
The second term $j_{t}$ is caused by the triplet component of SCDOS and vanishes in the equilibrium $V_{b\uparrow}=0$.
As it is seen in panel (b) of Fig.~\ref{SCDOS_fig}, the triplet part of SCDOS $N_{j,t}$ is an even function of 
energy and has finite value in the subgap energy region. So, multiplying it by the distribution function
\begin{equation}
\varphi_z^{(0)}+\tilde \varphi_z^{(0)}\approx({\rm sgn}[\varepsilon-eV_{b\uparrow}]-{\rm sgn}[\varepsilon
+eV_{b\uparrow}])/2
\label{distrib_tunnel_triplet}
\enspace ,
\end{equation}
one obtains current contribution $j_t$. The absolute value of this contribution
is roughly proportional to $V_{b\uparrow}$ for small enough values of this parameter and 
increases sharply when $V_{b\uparrow}$ approaches $|\Delta|$. This behavior is a consequence of two facts:
(i) distribution function (\ref{distrib_tunnel_triplet}) is a constant within energy interval 
$[-eV_{b\uparrow},eV_{b\uparrow}]$ and vanishes outside it, (ii) the triplet part of SCDOS has a particular shape
shown in panel (b) of Fig.~\ref{SCDOS_fig}. 

\begin{figure}[!tbh]
   \begin{minipage}[b]{\linewidth}
     \centerline{\includegraphics[clip=true,width=2.7in]{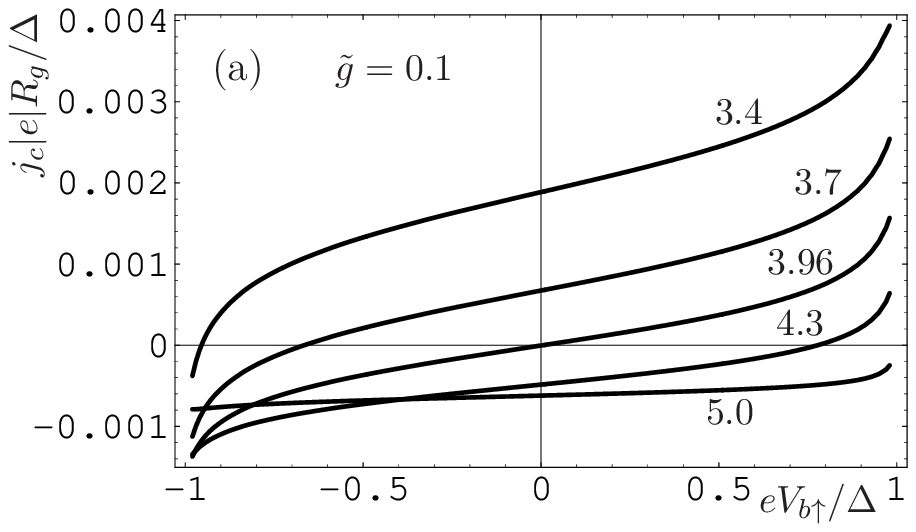}}
     \end{minipage}\hfill
    \begin{minipage}[b]{\linewidth}
   \centerline{\includegraphics[clip=true,width=2.64in]{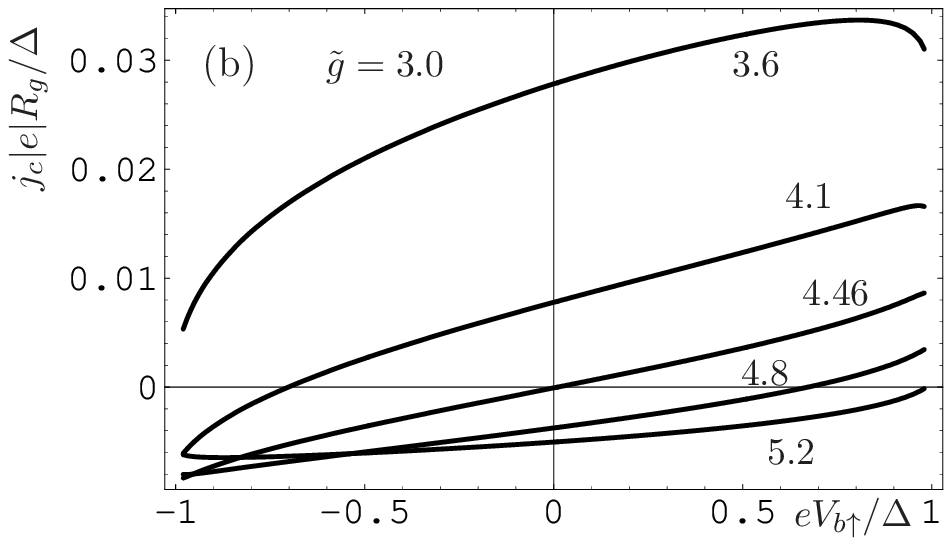}}
  \end{minipage}
   \caption{The critical Josephson current as a function of $e{V_b}_\uparrow/\Delta$. 
The different curves correspond to different lengths $d$ of the junction, which are 
measured in units of $\xi_F$.}
\label{current}
\end{figure}

\begin{figure}[!tbh]
   \begin{minipage}[b]{\linewidth}
     \centerline{\includegraphics[clip=true,width=2.7in]{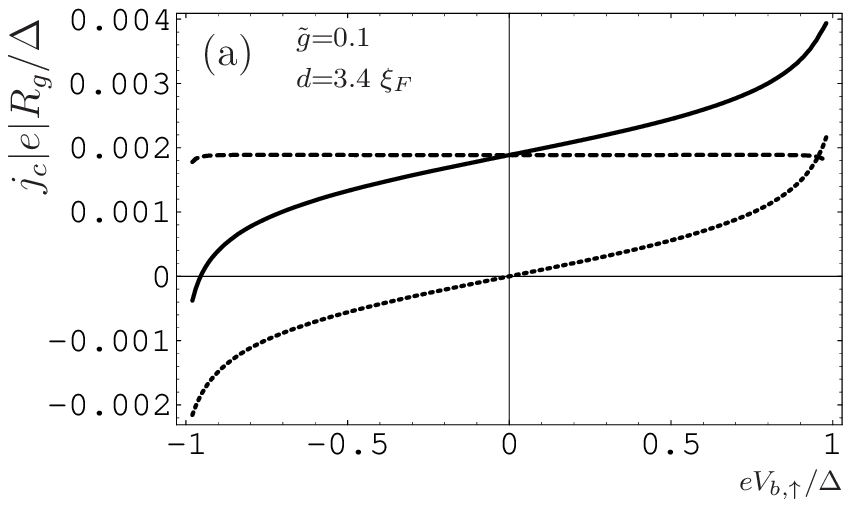}}
     \end{minipage}\hfill
    \begin{minipage}[b]{\linewidth}
   \centerline{\includegraphics[clip=true,width=2.64in]{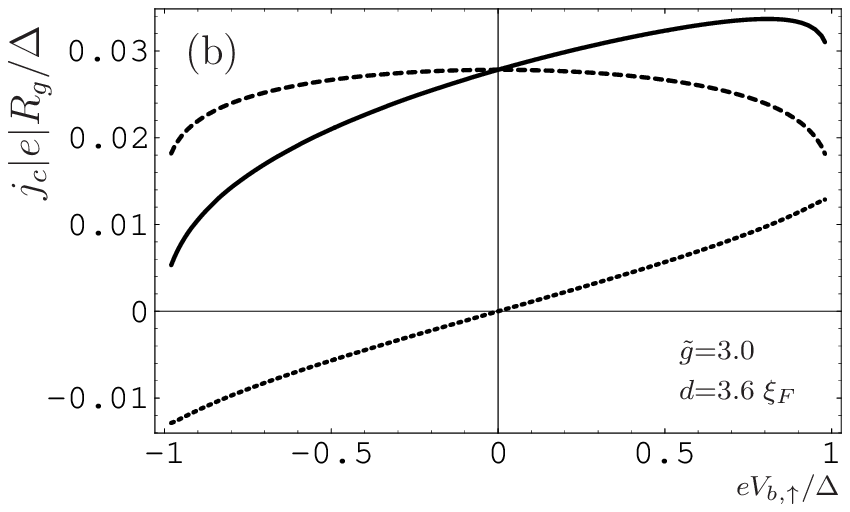}}
  \end{minipage}
   \caption{Separate plots of the current contributions $j_{s,c}$ (dashed line) and $j_{t,c}$ (dotted line)
together with the total critical current $j_{s,c}+j_{t,c}$ (solid line) as a function of $eV_{b\uparrow}/\Delta$. 
Panel (a) represents the low-transparency case and panel (b) corresponds to high-transparency case.}
\label{separate_currents}
\end{figure}

To calculate the Josephson current for the case of arbitrary transparency of SF interfaces one needs to solve Eq.~(\ref{K})
numerically and make use of Eq.~(\ref{current_general}). The resulting curves as functions of $eV_{b\uparrow}$
are plotted in Fig.~\ref{current}.
Panel (a) shows the current for low enough dimensionless conductance of SF interface $\tilde g=0.1$, while 
panel (b) represents the case of highly transparent interface $\tilde g=3.0$. Different curves correspond
to different lengths $d$ of the junction. In dependence on ${V_b}_\uparrow$ the current can be enhanced or reduced
with respect to its value at ${V_b}_\uparrow=0$. If the length of the equilibrium junction is not far from the
$0$-$\pi$ transition, then small enough voltage can switch between the states. Separate plots of the current contributions $j_s$ and $j_t$ are 
represented in panel (a) of Fig.~\ref{separate_currents} for the low-transparency junction with $\tilde g=0.1$
and in panel (b) for the high-transparency junction with $\tilde g=3.0$. It is seen, that for
low-transparency junction the current and its separate contributions $j_s$ and $j_t$ behave just as described
by the tunnel limit discussed above. While the tunnel limit qualitatively captures the essential physics
for the high-transparency junction as well, there are some new features, which are discussed below.

The singlet part of SCDOS for the case of high-transparency junction is plotted in panel (c) of Fig.~\ref{SCDOS_fig}.
It is clearly seen that as distinct from the limit of tunnel junction, it is not only concentrated
around superconducting gap edges, but is finite in the whole subgap region. It results in the sensitivity of
$j_s$ to the fact that the distribution function $\varphi^{(0)}_0+\tilde \varphi^{(0)}_0$ is nonequilibrium.
In other words, upon increasing of the SF-interface conductance the mechanism of the current controlling 
based on the redistribution of the quasiparticles between energy levels starts to contribute. 
It is seen from panel (b) of Fig.~\ref{separate_currents} that the difference 
$j_{s,c}(V_{b\uparrow})-j_{s,c}(V_{b\uparrow}=0)$ considerably grows when $|eV_{b\uparrow}|$ approaches $|\Delta|$.
It is worth to note that the absolute value of $j_{s,c}$ is always reduced by this mechanism. This fact
leads to the reduction of the absolute value of the high-transparency total supercurrent (see Fig.~\ref{current}(b))
at $eV_{b\uparrow} \to \Delta$. This behavior should be compared to the tunnel limit, where 
the dependence of $j_{s,c}$ on $V_{b\uparrow}$ is negligible and absolute value of the total supercurrent 
only increases for $eV_{b\uparrow} \to \Delta$ due to growing contribution of $j_{t,c}$.

\begin{figure}[!tbh]
  \centerline{\includegraphics[clip=true,width=3.5in]{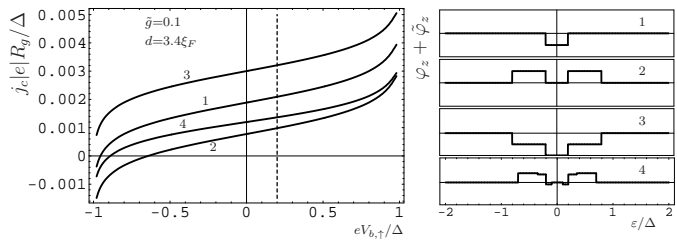}}
   \caption{Dependence of the critical current on the particular shape of the distribution function. 
The right panel represents four different examples of the distribution function vector part. The curves
marked by the numbers from 1 to 4 correspond to the following sets of parameters in Eq.~(\ref{phi_N}):
1.~$g_{t\sigma} \to 0$, $V_{b\downarrow}=0$; 2.~$g_{t\uparrow},g_{b\downarrow} \to 0$, $eV_{t\downarrow}=0.8\Delta$;
3.~$g_{t\uparrow},g_{b\downarrow} \to 0$, $eV_{t\downarrow}=-0.8\Delta$ and 
4.~$g_{t\uparrow}=0.1\sigma_F/d_y$, $g_{t\downarrow}=4\sigma_F/d_y$, $g_{b\uparrow}=3\sigma_F/d_y$, 
$g_{b\downarrow}=0.2\sigma_F/d_y$, $V_{b\downarrow}=0.5V_{b\uparrow}$, $eV_{t\uparrow}=0.35\Delta$, 
$eV_{t\downarrow}=0.7\Delta$. 
The respective critical current plots in dependence on $eV_{b\uparrow}$,
calculated for the same sets 1-4 of the distribution function parameters are shown in the left panel.
The dashed vertical line marks the position $eV_{b\uparrow}=0.2\Delta$, for which the distribution functions
are calculated.}   
\label{distrib_dependence}
\end{figure}

Now we discuss the dependence of the obtained results to the particular shape of the distribution function.
It is illustrated in Fig.~\ref{distrib_dependence}. Four different examples of the distribution function
z-component $\varphi_z^{(0)}+\tilde \varphi_z^{(0)}$, satisfying Eq.~(\ref{phi_N}) are shown in the right panel.
Due to symmetry relations this combination is always an even function of energy. The scalar
part of the distribution function is not plotted in the figure because it does not influence the supercurrent
in the tunnel limit. The corresponding
plots of the critical current in the low-transparency limit are represented in the left panel. It can be concluded
that, while the quantitative value of the critical current depends on the particular choice of the distribution
function, this choice has no qualitative effect on the supercurrent behavior, as it was already pointed before.                   

In conclusion, we have studied the effects of nonequilibrium spin-dependent electron distribution in a 
weakly ferromagnetic interlayer on the Josephson current through SFS junction. It is shown that the
nonequilibrium spin-dependent electron distribution gives rise to the supercurrent carried by the triplet
component of SCDOS. Depending on voltage, controlling the particular form of spin-dependent 
nonequilibrium in the interlayer, this additional current can enhance or reduce the usual current of the singlet
component and also switch the junction between $0$- and $\pi$-states. 

We gratefully acknowledge discussions with V.V. Ryazanov, A.S. Melnikov and M.A. Silaev. The
support by RFBR Grant 09-02-00779 and the programs of Physical Science Division
of RAS is acknowledged. A.M.B. was also supported by the Russian
Science Support Foundation.


\end{document}